\documentclass[12pt]{iopart}
\usepackage{graphicx}
\usepackage{color}

\begin{document}

\title[Analysis of adiabatic transfer...]{Analysis of adiabatic transfer in cavity QED} 

\author{Joyee Ghosh\footnote{Present address: ICFO -- Institut de Ciencies Fotoniques, Mediterranean Technology Park, 08860 Castelldefels (Barcelona), Spain}, R Ghosh and Deepak Kumar}

\ead{rghosh.jnu@gmail.com}

\address{School of Physical Sciences, Jawaharlal Nehru University, New Delhi 110 067, India}

\date{12 July 2010}

\begin{abstract}
A three-level atom in a $\Lambda$ configuration trapped in an optical cavity forms a basic unit in a number of proposed protocols for quantum information processing. This system allows for efficient storage of cavity photons into long-lived atomic excitations, and their retrieval with high fidelity, in an adiabatic transfer process through the `dark state' by a slow variation of the control laser intensity. We study the full quantum mechanics of this transfer process with a view to examining the non-adiabatic effects arising from inevitable excitations of the system to states involving the upper level of the $\Lambda$, which is radiative. We find that the fidelity of storage is better, the stronger the control field and the slower the rate of its switching off. On the contrary, unlike the adiabatic notion, retrieval is better with faster rates of switching on of an optimal control field. Also, for retrieval, the behavior with dissipation is non-monotonic. These results lend themselves to experimental tests. Our exact computations, when applied to slow variations of the control intensity for strong atom-photon couplings, are in very good agreement with Berry's superadiabatic transfer results without dissipation.
\end{abstract} 

\pacs{03.67.-a, 42.50.Ex, 42.50.Gy, 42.50.Ct} 

\noindent{\it Keywords\/}: cavity QED, adiabatic transfer, storage and retrieval, quantum information

\submitto{\jpb}

\maketitle

\section{Introduction}
\label{intro}

In the context of current efforts to build quantum networks \cite{kimble08}, a promising way to transfer quantum states reliably in the network is through the coupling of single photons and atoms in the setting of cavity quantum electrodynamics (QED) \cite{miller05}. For practical applications, the coupling between a single atom located in an optical cavity and a single intercavity photon should be strong. The strong coupling condition requires that $g_0/2 \gg \Gamma, \kappa$, where $g_0$ is the one-photon Rabi frequency, $\Gamma$ is the atomic decay rate to modes other than the cavity mode, and $\kappa$ is the decay rate of the cavity mode itself. This may be achieved using high-finesse optical cavities, with an extreme reduction in the cavity volume, and using atomic transitions with a large dipole moment.

Using the cavity QED techniques, schemes for a number of operations of direct relevance to quantum information processing have been proposed, one of them being generation of single photons `on demand'. Through strong coupling of a mode of the cavity field to an atomic transition, which is resonantly driven by the input single-photon pulse, an external control field of Rabi frequency $\Omega_C(t)$ transfers one photon in the cavity mode to a long-lived atomic memory, which can then be released at will to free space through the cavity output mirror, leading to an output single-photon pulse as a collimated beam. The temporal structure (both amplitude and phase) of the resultant `flying photon' can be tailored by way of the control field $\Omega_C(t)$ \cite{cirac97,duan04}, with the spatial structure of the wave-packet being set by the cavity mode. 

The basic scheme (see figure 1) involves a three-level atom in a $\Lambda$-configuration with an excited state $|a \rangle$ and two lower states $|b \rangle$ and $|c \rangle$. An optical cavity mode is strongly and coherently coupled to the atom on the $|b \rangle \leftrightarrow |a \rangle$ transition with rate $g_0$, and a strong classical field $\Omega_C(t)$ drives the $|c\rangle \leftrightarrow |a \rangle$ transition. 
Denoting by $|x,n \rangle$ a state in which the atom is in state $|x \rangle$ and there are $n$ photons in the cavity mode, reversible transfer of a state between light and a single trapped atom can be achieved through the mappings $|b,1 \rangle \leftrightarrow |c,0 \rangle $ for the coherent absorption and emission of single photons by a procedure involving the `dark state'. 
\begin{figure}
\centering
\resizebox{0.75\columnwidth}{!}{%
    \includegraphics{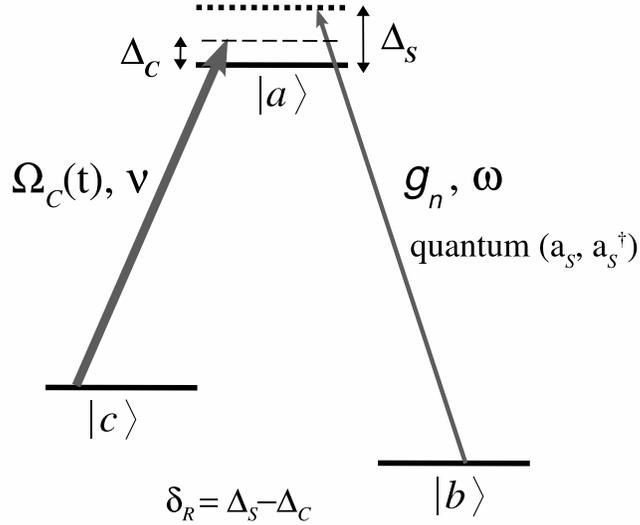}
}
\caption{Three-level $\Lambda$ scheme}
    \label{lambda}
\end{figure}
The atom-cavity system coupled to a classical control field $\Omega_C(t)$ has an instantaneous eigenstate $|D \rangle$, given by
\begin{equation}
|D \rangle = \cos \theta (t) \vert b,1 \rangle - \sin\theta (t) \vert c,0 \rangle, \label{eq1} 
\end{equation}
where
\begin{equation}
\tan\theta (t) = \frac{g_0}{\Omega_C(t)} . \label{eq2}
\end{equation}
(\ref{eq1}) is called the `dark state' \cite{bergRMP98} as it does not have any component involving the excited level $|a \rangle$, which can radiate. If initially the system is prepared in the state $|c, 0 \rangle$ with $\Omega_C(t = 0) = 0$, then a sufficiently slow increase of $\Omega_C(t \rightarrow \infty)$ to $\Omega_0 \gg g_0$ leads the state to move adiabatically to $|b, 1 \rangle$. On the other hand, if initially the system is in state $|b, 1 \rangle$ with $\Omega_C(t = 0) = \Omega_0$, then decreasing $\Omega_C(t)$ adiabatically to zero leads the system to the state $|c, 0 \rangle$. The advantages of the dark state protocol are: (a) it minimizes the dissipative effects generic to two-level systems, and (b) it is reversible, i.e., a photon that is emitted from a system A is efficiently transferred to another system B by applying the time-reversed (and suitably delayed) control field $\Omega_C(t)$ to system B. 

Cirac {\it et al.} \cite{cirac97} proposed a quantum network with nodes consisting of cavities each containing a three-level atom. The entanglement between atom-cavity states at each node is generated using the protocol described above. Similar protocols involving adiabatic transfers using three-level atoms, either trapped in or passing through electromagnetic cavities, have also been proposed for other purposes. Parkins {\it et al.} \cite{parkins93} were the first to propose this kind of protocol to generate Fock states and other nonclassical states of the cavity mode. Pellizzari {\it et al.} \cite{zoller95} proposed implementation of a two-bit quantum gate by putting two three-level atoms inside the cavity. 
The first experimental attempt of such a reversible mapping of a field to and from an atomic state has been made \cite{boozer07} by using a single trapped cesium atom. In this example, $|c \rangle$ and $|b \rangle$ represent internal states of the atom with long-lived coherence, namely, the hyperfine states in the 6S$_{1/2}$, F = 3 and F = 4 manifolds of atomic Cs, and $|a \rangle$ corresponds to 6P$_{3/2}$, F = 3. Intrinsically reversible and controlled single-photon sources have been demonstrated \cite{rempe99,rempe00,rempe02,rempe07} using the dark state based on stimulated Raman adiabatic passage between two ground states of a single atom strongly coupled to a single mode of a high-finesse optical cavity. 

The above kind of adiabatic transfer has also been used with {\it ensembles} of three-level atoms in a pencil geometry, which lead to the well known phenomena of electromagnetically induced transparency (EIT) and slow light \cite{kk88,harris89,boller91,lukin00,liu01,expts,jg08}. With atomic ensembles, the light pulses can be stored as collective atomic polaritons and recovered with high fidelity again by manipulating a control laser adiabatically \cite{fleischhauer02}. Duan {\it et al.} \cite{dlcz01} have proposed long-distance quantum communication and entanglement using nodes of atomic ensembles. Very recently, {\it single-atom} EIT condition through a coherent dark state has been achieved in a high-finesse optical cavity \cite{rempe10, meschede10}. The single atom effectively acts as a quantum optical transistor, coherently controlling the transmission of light through the cavity. 

Though the various procedures outlined above have different physical features, underlying all of them is the assumption of adiabatic transfer through the dark state of a single three-level atom in the $\Lambda$-configuration. The adiabatic condition is taken to be fulfilled if the evolution time is significantly longer than the inverse of the frequency gap between the dark state and the other eigenstates of the Hamiltonian. But it is of utmost interest to understand the precise conditions, the fidelity and the experimental limitations of this procedure which allows transfer of quantum state information from the photon to an atom and vice versa. 

This question has received a lot of attention in several physical contexts. In general, the superadiabatic transfer procedure given by Berry \cite{berry90} allows one to treat the non-adiabatic corrections in the quantum evolution for slow rates of change of the control laser in the absence of dissipation. For cavity-QED with a trapped atom, Duan, Kuzmich and Kimble \cite{dkk2003} have presented a detailed analysis of schemes which require photon transfer between cavity mode and the mode of the external channel. Here the non-adiabatic analysis requires inclusion of a large number of states, and the analysis has been carried out numerically. A drawback of this analysis is that the dissipative terms are put in the Schr\"{o}dinger equation for the amplitudes. This may suppress some important non-adiabatic elements of the quantum transfer process, and the issue needs to be examined, if possible, using a better formalism. The same problem has also been analyzed by Yao {\it et al.} \cite{yls2005}, but in their analysis, the coupling to the external channel has been treated in the Wigner-Weisskopf approximation, which may not hold for all the procedures of interest. 

We wish to address here the exact problem of the transfer dynamics in the context of a single atom in an electromagnetic cavity, for arbitrary rates of variation of the control laser and with the inclusion of dissipation in the form of spontaneous emission from the upper level. For a three-level atom in the $\Lambda$ configuration interacting with a single cavity mode, one can carry out a thorough non-adiabatic analysis as the Hilbert space is small. As has been seen in \cite{zoller95}, the effect of spontaneous photons from the excited atomic state on quantum computation is much more destructive than cavity decay. This is because after a cavity photon emission, the system is still in a dark state, which is not necessarily true for spontaneous emission. Thus the atomic excited states can be populated, which gives rise to further spontaneous photons. Furthermore, we work in the strong-coupling regime where $g_0 \gg \kappa$. In this regime we neglect the decay of the cavity photon. Thus our analysis is not directly applicable to cases where photon transfer to external channel is of comparable time scale.

The paper is organized as follows. In section 2, we first study the problem of an atomic system in a $\Lambda$ configuration interacting with photons in a cavity mode and a time-dependent control field, ignoring any dissipative effects. Our simple considerations show that the non-adiabatic effects are large when $\Omega_C(t)$ is small. Since during storage as well as retrieval, $\Omega_C(t)$ is made zero, the non-adiabatic effects are unavoidable, and the problem cannot be treated perturbatively. Accordingly, we study the problem numerically.

Next, in section 3, we analyze the effect of dissipation by allowing for the possibility of spontaneous emission, which is inevitable in any realistic system. Since spontaneous decay sends the system to the lower levels, one might expect that dissipation would mitigate the non-adiabatic effects. We study the impact of dissipation using a method in which the quantum evolution of the wavefunction is interrupted by spontaneous transitions of the state to the lower levels $\vert b \rangle$ and $\vert c \rangle$. Spontaneous decays are governed by a stochastic Poissonian process with a rate which we take to be the width of the level $\vert a \rangle$. Though dissipation in quantum systems is naturally incorporated in the density matrix formalism, we believe that the wavefunction treatment is a fair approximation, and it has two distinct advantages. First, the physical picture of the system evolution and the role of dissipation is rather transparent. Second, there is a numerical simplification of having to solve only three coupled time-dependent differential equations.

In section 4, we present our results on the dynamics of the fidelity of the storage and the retrieval process. Different rates of variation of the control field at different signal strengths are explored, each in the absence and presence of dissipation. We compare our results in the absence of dissipation to the superadiabatic theory due to Berry \cite{berry90}. Finally, in section 5, we present our conclusions. In Appendix A, we compare the wavefunction approach with the density matrix approach and point out the nature of the approximation made in our analysis. 

\section{Formulation with a single isolated atom}
\label{single-atom}

We now consider the $\Lambda$ atomic system as shown in figure 1, which interacts with the signal photon in the cavity mode and the control laser field. The Hamiltonian of the isolated atom-signal field system is $\textbf{H}\mbox{\boldmath $_0$} +\textbf{H}\mbox{$_{\mathrm{I}}$}$, with 
\begin{equation}
\textbf{H}\mbox{\boldmath $_0$} = \hbar \omega \left( \mathbf{a}_S^{\dagger} \mathbf{a}_S + \frac{1}{2} \right)  + \displaystyle\sum_{x} E_x \vert x \rangle \langle x \vert , \label{eq5}
\end{equation}
where $\mathbf{a}_S$ denotes the annihilation operator for the particular cavity mode of frequency $\omega$, which couples to the transition between levels $|a \rangle$ and $|b \rangle$, and $x = a, b, c$. In the rotating-wave approximation, the time-dependent interaction Hamiltonian of the fields with the atomic system (in one dimension) is
\begin{eqnarray}
\textbf{H}\mbox{$_{\mathrm{I}}$} = && \hbar g_0 \left[ \vert a \rangle \langle b \vert \mathbf{a}_S + \mathbf{a}_S^{\dagger} \vert b \rangle \langle a \vert \right] - \hbar \Big[ \Omega_C (t) e^{-i \nu t} \vert a \rangle \langle c \vert \nonumber \\ && + \Omega_C^* (t) e^{i \nu t} \vert c \rangle \langle a \vert \Big] , \label{eq6}
\end{eqnarray}
where $\nu$ is the frequency of the control field, and $g_0$, the Rabi frequency for the cavity photon is given by 
\begin{equation}
\hbar g_0 = \mu_{ab} \sqrt{\frac{2 \hbar \omega}{\epsilon_0 V_0}} , \label{eq7}
\end{equation}
with $\mu_{ab}$ being the dipole moment between levels $|a \rangle$ and $|b \rangle$, $V_0$ the cavity volume, and $\epsilon_0$ the permittivity of vacuum. It is assumed that only transitions $\vert a \rangle \rightarrow \vert b \rangle$ and $\vert a \rangle \rightarrow \vert c \rangle$ are dipole-allowed. The control field with a Rabi frequency $\Omega_C$ is treated classically. 

Considering an $n$-photon quantum field, the wavefunction of the system of one atom+field can be expressed in general as
\begin{eqnarray}
\vert \Psi (t) \rangle = && \displaystyle\sum_{n} \Big[ A_{n}(t) e^{-i \omega_{an} t} \vert a,n \rangle + B_{n}(t) e^{-i \omega_{bn} t} \vert b,n \rangle \nonumber \\ && + C_{n}(t) e^{-i \omega_{cn} t} \vert c,n \rangle \Big]. \label{eq8}
\end{eqnarray}
Here 
\begin{equation}
 \hbar \omega_{xn} = E_x + \left( n+\frac{1}{2} \right) \hbar \omega . \label{eq9}
\end{equation}
For this one-atom case, a closed set of equations of motion for the coefficients $A_n$, $B_n$ and $C_n$ are obtained. For further analysis, it is convenient to work with the vector {\bf X} with components
\begin{eqnarray}
 a_n = A_n, ~~ b_n = e^{-i \Delta_S t} B_{n+1}, ~~ c_n = e^{-i \Delta_C t} C_n , \label{eq11}
\end{eqnarray}
where $\Delta_S \equiv \omega - \omega_{ab}$ and $\Delta_C \equiv \nu - \omega_{ac}$ denote, respectively, the detunings of the two fields. Then {\bf X} obeys the equation 
\begin{equation}
 i \frac{d \textbf{X}}{dt} = \textbf{H}(t) ~\textbf{X}(t), \label{eq15}
\end{equation}
where
\begin{eqnarray}
 \textbf{H}(t)=\left( \begin{array}{ccc}
                0 & g_n & -\Omega_{C}(t) \\
		g_n & \Delta_S & 0 \\
		-\Omega_C^* (t) & 0 & \Delta_C
               \end{array} \right) , \label{eq16}
\end{eqnarray}
with $g_n = g_0 \sqrt{n+1}$ denoting the Rabi frequency of the $n$-photon signal field. 

We write the solution of the above equation in terms of the instantaneous eigenvalues and eigenvectors:
\begin{equation}
\textbf{H}(t) ~|u_k(t) \rangle = \hbar \lambda_k(t) ~|u_k(t) \rangle, ~~ k = 1, 2, 3 . \label{eq19}
\end{equation}
At two-photon resonance, i.e., with $\Delta_S = \Delta_C \equiv \Delta$, the instantaneous eigenvalues are
\begin{eqnarray}
\lambda_1 = \Delta, ~~ \lambda_2 = \frac{\Delta}{2} + \Omega_R(t), ~~ \lambda_3 = \frac{\Delta}{2} - \Omega_R(t), \label{eq20}
\end{eqnarray}
where 
\begin{eqnarray}
\Omega_R(t) &=& \frac{\sqrt{\Delta^{2} + 4 \Omega_{\mathrm{eff}}^2(t)}}{2}, \nonumber \\ 
\Omega_{\mathrm{eff}}(t) &=& \sqrt{g_n^2 + \vert \Omega_C (t) \vert ^2}. 
\end{eqnarray}
The instantaneous eigenvectors of the atom-field system are
\begin{eqnarray}
\vert u_1(t) \rangle &=& \cos{\theta(t)} e^{i \phi} ~\vert b,n+1 \rangle + \sin{\theta(t)} ~\vert c,n \rangle, \nonumber \\
\vert u_2(t) \rangle &=& \cos{\frac{\psi(t)}{2}} ~\vert a,n \rangle + \sin{\frac{\psi(t)}{2}} \Big[ \sin{\theta(t)} ~\vert b,n+1 \rangle \nonumber \\ && - \cos{\theta(t)} e^{-i \phi} ~\vert c,n \rangle \Big], \nonumber \\
\vert u_3(t) \rangle &=& - \sin{\frac{\psi(t)}{2}} ~\vert a,n \rangle + \cos{\frac{\psi(t)}{2}} \Big[ \sin{\theta(t)} ~\vert b,n+1 \rangle \nonumber \\ && - \cos{\theta(t)} e^{-i \phi} ~\vert c,n \rangle \Big]. \label{eq21}
\end{eqnarray}
Here $\phi$ is some arbitrary constant phase of the control field, and
\begin{eqnarray}
\tan{\theta(t)} &=& \frac{g_n}{\Omega_C(t)},  \label{eq22} \\
\tan{\psi(t)} &=& \frac{\Omega_{\mathrm{eff}}(t)}{\Delta/2}. \label{eq23}
\end{eqnarray}
The eigenstate $\vert u_1(t) \rangle$ is the `dark state'. Now we can expand the solution in terms of these eigenvectors as
\begin{equation}
\vert \Psi (t) \rangle = \displaystyle\sum_{k} D_k(t) e^{-i \int_{0}^{t} \lambda_k(t') dt'} \vert u_k(t) \rangle. \label{eq24}
\end{equation}
The time-dependent coefficients $D_m(t)$ obey the following equation:
\begin{eqnarray}
&&\frac{d D_m(t)}{d t} + D_m(t) \langle u_m(t) \vert \dot{u}_m(t) \rangle \nonumber \\ =&& - \displaystyle\sum_{k \neq m} D_k(t) \langle u_m(t) \vert \dot{u}_k(t) \rangle e^{-i \int_{0}^{t} (\lambda_k(t') - \lambda_m(t')) dt'} . \label{eq25}
\end{eqnarray}
By making a change of variable to
\begin{equation}
V_m(t) = D_m(t) e^{i \int_{0}^{t} \beta_m(t') dt'}, ~~~ i \beta_m(t') = \langle u_m(t) \vert \dot{u}_m(t) \rangle , \label{eq26}
\end{equation}
the evolution of $V_m(t)$ is obtained as
\begin{eqnarray}
\frac{d V_m(t)}{d t} = - \displaystyle\sum_{k \neq m} V_k(t) \langle u_m(t) \vert \dot{u}_k(t) \rangle e^{-i \int_{0}^{t} \lambda'_{km}(t') dt'}, \label{eq27}
\end{eqnarray}
where $\lambda'_{km} = \lambda_{km} + \beta_{km}$, $\lambda_{km} = \lambda_k - \lambda_m $ and $\beta_{km} = \beta_k - \beta_m $. Note that the assumption of adiabaticity implies that the coefficients $D_m$s or $V_m$s are independent of time. The time-variation of these coefficients is governed by the terms $\langle u_m(t) \vert \dot{u}_k(t) \rangle$, which essentially give rise to non-adiabatic effects. From the eigenvectors in Eqs.\,($\ref{eq21}$), we derive the following:
\begin{eqnarray}
\langle u_1 \vert \dot{u}_2 \rangle &=& -\langle u_2 \vert \dot{u}_1 \rangle^* = \dot{\theta}(t) ~\sin{\frac{\psi(t)}{2}} ~e^{-i \phi}, \nonumber \\
\langle u_1 \vert \dot{u}_3 \rangle &=& -\langle u_3 \vert \dot{u}_1 \rangle^* = \dot{\theta}(t) ~\cos{\frac{\psi(t)}{2}} ~e^{-i \phi}, \nonumber \\
\langle u_2 \vert \dot{u}_3 \rangle &=& -\langle u_3 \vert \dot{u}_2 \rangle^* = -\frac{\dot{\psi}(t)}{2} \label{eq33}, 
\end{eqnarray}
where
\begin{eqnarray}
\dot{\theta}(t) &=& -\frac{g_n}{g_n^2 + {\vert \Omega_C(t) \vert}^2} ~\frac{d\Omega_C}{dt}, \label{eq34} \\
\dot{\psi}(t) &=& \frac{4 \Delta \Omega_C(t)}{\sqrt{g_n^2 + {\vert \Omega_C(t) \vert}^2} ~\left[ \Delta^2 + 4(g_n^2 + {\vert \Omega_C(t) \vert}^2) \right]} \nonumber \\ && \times \frac{d\Omega_C}{dt}. \label{eq35}
\end{eqnarray}
All $\beta_k$s are zero, and therefore $\lambda'_{km} = \lambda_{km}$.
 
Then, finally we can express the coefficients $A_n(t)$, $B_{n+1}(t)$, $C_n(t)$ of our initial basis of the bare states in (\ref{eq8}) in terms of these solutions as
\begin{eqnarray} 
A_n(t) &=& \Big( \cos\frac{\psi(t)}{2} V_2(t) e^{-i \int_{0}^{t} \lambda'_{2}(t') dt'} \nonumber \\ && - \sin\frac{\psi(t)}{2} V_3(t) e^{-i \int_{0}^{t} \lambda'_{3}(t') dt'} \Big), \label{eq30} \\
B_{n+1}(t) &=& V_1(t) e^{-i \int_{0}^{t} \lambda'_{1}(t') dt'} \cos{\theta(t)} e^{i\phi} + \sin{\theta(t)} \nonumber \\ && \times \Big( \sin\frac{\psi(t)}{2} V_2(t) e^{-i \int_{0}^{t} \lambda'_{2}(t') dt'} \nonumber \\ && + \cos\frac{\psi(t)}{2} V_3(t) e^{-i \int_{0}^{t} \lambda'_{3}(t') dt'} \Big) , \label{eq31} \\
C_n(t) &=& V_1(t) e^{-i \int_{0}^{t} \lambda'_{1}(t') dt'} \sin{\theta(t)} - \cos{\theta(t)} \nonumber \\ && \times e^{-i\phi} \Big( \sin\frac{\psi(t)}{2} V_2(t) e^{-i \int_{0}^{t} \lambda'_{2}(t') dt'} \nonumber \\ && + \cos\frac{\psi(t)}{2} V_3(t) e^{-i \int_{0}^{t} \lambda'_{3}(t') dt'} \Big) . \label{eq32}  
\end{eqnarray}

From the above equations, it is quite evident that the non-adiabatic perturbation, which is proportional to $\dot{\theta}$, becomes large in the storage and retrieval process as $\Omega_C(t)$ becomes small. Thus we solve Eqs.\,(\ref{eq27}) for $V_m$s numerically. For simplicity, we take the phase of the control field, $\phi=0$, and the optical detuning, $\Delta=0$. The latter would imply that $\psi = \frac{\pi}{2}$ and $\dot{\psi} = 0$, leading to considerable numerical simplification.

It is interesting to place the significance of our analysis in the context of a known general result on adiabaticity for three-level systems obtained by Oreg \textit{et al.} \cite{ohe84}. These authors analyzed the density-matrix equations as SU(3) rotations of an 8-component vector $\vec{S}$ constructed out of 8 independent components of the density matrix. It was shown that the stationary sector of the solution consists of two vectors $\vec{\Gamma}_1$ and $\vec{\Gamma}_2$, which are obtained from the Hamiltonian parameters. This implies that $\dot{\vec{S}} = 0$, when $\vec{S}$ is any linear combination of $\vec{\Gamma}_1$ and $\vec{\Gamma}_2$. Such an $\vec{S}$ follows the subspace of $\vec{\Gamma}_1$ and $\vec{\Gamma}_2$ adiabatically when the Hamiltonian parameters change in time, allowing for an adiabatic transfer of the quantum state within the subspace. A qualitative measure of non-adiabatic effects is provided by the angle $\chi$, which is between $\vec{S}$ and its projection onto the subspace,
\[ \cos \chi = \left( D_1^2 + D_2^2 \right)^{1/2} , \]
where $D_i = \vec{S} \cdot \vec{\Gamma}_i$. When $\chi \approx 0$, the adiabatic following of $\vec{S}$ with the subspace is good. Following Oreg \textit{et al.} \cite{ohe84}, we can easily obtain $\chi$ for the dark state $\vert u_1(t) \rangle$. The vectors $\vec{S}$ and $\vec{\Gamma}_1$ for the above Hamiltonian are seen to be
\[ \vec{S} = \left( 0, 0, -\sin (2\theta), 0, 0, 0, \cos^2 \theta, \frac{1}{\sqrt{3}}(1 - 3 \sin^2 \theta ) \right) , \]
\begin{eqnarray}
\vec{\Gamma}_1 = \frac{1}{\sqrt{g_n^2 + \Omega_C^2 + 4 \Delta^2/3}} \Big( && -g_n, -\Omega_C, 0, 0, 0, 0, \nonumber \\ && \Delta, -\Delta/\sqrt{3} \Big) . \nonumber
\end{eqnarray}
The calculation is straightforward and we just quote the result for $\Delta = 0$: $D_1 = D_2 = 0$ and $\chi = \pi /2$, thus making non-adiabatic effects rather strong.

\subsection{Fidelity of storage and retrieval}

We first consider the {\it storage} of a single photon from the cavity to the atomic memory. The initial state in this situation has $\Omega_C(0) = \Omega_0$, and the atom-signal system is in the dark state (with zero eigenvalue),
\begin{equation}
|\Psi(0) \rangle = \cos\theta(0) \vert b,1 \rangle + \sin\theta(0) \vert c,0 \rangle = \vert u_1(0) \rangle , \label{eq52} 
\end{equation}
where
\begin{equation}
\theta(0) = {\mathrm{tan}}^{-1} ~\left( \frac{g_0}{\Omega_0} \right) \label{eq53}.
\end{equation}
The signal pulse is to be stored by making the control field $\Omega_C (t)$ zero. A suitable form \cite{fleischhauer02} for the control pulse is
\begin{equation}
\Omega_C (t) = \Omega_0 \left[ 1-\tanh(rt) \right]. \label{eq50}
\end{equation}
The adiabatic evolution of the above state leads to just $\vert u_1(t) \rangle$ at time $t$. After a lapse of time of the order $3/r$, we expect the wavefunction to evolve to $|c,0 \rangle$ as $\Omega_C(t) \rightarrow 0$ i.e., $\theta (t) \rightarrow \frac{\pi}{2}$.

As a measure of any departure of our solution $|\Psi(t) \rangle$ from the adiabatic answer, we compute the fidelity $F(t)$ of the process given by 
\begin{equation}
F(t) = |\langle u_1(t)|\Psi(t) \rangle | \label{eqfidelity},
\end{equation}
for different values of $g_0$ and $r$. We also compute $|A_0(t)|^2$, $|B_1(t)|^2$ and $|C_0(t)|^2$ to portray the actual evolution of the state. 

For the {\it retrieval} of the photon from atomic memory, we need to increase the control field from zero to $\Omega_0$. For this, we take the control pulse to be of the form \cite{fleischhauer02}
\begin{equation}
\Omega_C (t) = \Omega_0 \tanh(rt) \label{controlret}.
\end{equation}
The initial state of the system is the dark state (\ref{eq52}), now with $\theta (0) = \frac{\pi}{2}$. Again, the fidelity (\ref{eqfidelity}) of the process records the deviation from the adiabatic evolution. 

\section{Wavefunction formulation in the presence of dissipation}

There are standard ways of incorporating dissipation in quantum systems. A comprehensive account of those which are of particular use in quantum optics can be found in the textbook by Scully and Zubairy \cite{szbook}. For a three-level system, the density matrix equations incorporating dissipation have been investigated in the literature in a somewhat different context \cite{java,gsa}. Here we adopt an approach which we believe to be quite transparent from a physical point of view, as supported by our results in the next section. We work directly with the wavefunction \cite{dalibard92}, and this formulation can be regarded as an approximation to the full set of density matrix equations. We present a discussion of this approximation with regard to density-matrix treatment in Appendix A. Here we remark that our approach is similar in spirit to the formulation of Dalibard \textit{et al.} and of Barchielli and Belavkin for a continuously measured system \cite{bb91,dalibard92}, and has also been used in a similar context for $\Lambda$-systems in interaction with cavity fields \cite{parkins93,cirac97}.

In this approach \cite{cb71}, the unitary evolution of the system of a three-level atom interacting with the signal and control fields is interrupted by spontaneous decays. We assume that in the presence of spontaneous decay, which occurs over a negligible time, the system collapses to either level $\vert b \rangle$ or level $\vert c \rangle$ with equal probabilities. The decays occur in time according to a Poissonian distribution. To write down the wavefunction in this model, we use the following notation. $U(t)$ denotes the evolution operator for the isolated system, which is computed in the previous section through the computation of $V_m(t)$s in (\ref{eq27}). The probability that a spontaneous decay occurs in the time interval $dt$ is denoted by $\Gamma dt$. The probability $P(t)$ that a decay has not occurred for time $t$, after preparation of the system at $t=0$, is $e^{-\Gamma t}$. The operators that cause spontaneous decays to states $\vert b \rangle$ and $\vert c \rangle$ are, respectively, denoted by $\zeta_b$ and $\zeta_c$. In writing down the wavefunction at time $t$, we have to include the possibilities of $0, 1, 2, \ldots , l, \ldots$ decays, with each of these weighted by the probability distribution mentioned above, and thus 
\begin{equation}
|\Psi(t) \rangle = \frac{1}{Z(t)} \sum_{l=0}^{\infty}Q_l |\Psi(0) \rangle , \label{series}
\end{equation}
where $Q_l$s denote the possibility in which $l$ spontaneous decays have occurred over the interval $t$. These are given as
\begin{eqnarray}
Q_0 &=& e^{-\Gamma t} U(t), \nonumber \\
Q_1 &=& \int_{0}^{t} dt_1 ~e^{-\Gamma(t-t_1)} U(t-t_1) ~\Gamma ~\zeta ~e^{-\Gamma t_1} U(t_1),  \nonumber \\
\vdots \nonumber \\
Q_l &=& \int_{0}^{t} \int_{0}^{t_1} \int_{0}^{t_2} ~\cdots \int_{0}^{t_{l-1}} dt_1 ~dt_2... ~dt_l ~e^{-\Gamma(t-t_1)} \nonumber \\ && \times U(t-t_1) ~\Gamma ~\zeta ~e^{-\Gamma (t_1 - t_2)} U(t_1-t_2) \Gamma ~\zeta \ldots \nonumber \\ && \times e^{-\Gamma (t_{l-1} - t_l)} U(t_{l-1}-t_l) ~\Gamma ~\zeta e^{-\Gamma t_l} U(t_l) , \label{eq69}
\end{eqnarray}
where $\zeta$ denotes either $\zeta_b$ or $\zeta_c$. We shall also average over these stochastic histories by assuming the decays to be independent. In (\ref{series}), $Z(t) = \sqrt{\langle\Psi(t)|\Psi(t) \rangle}$ is the normalization of the wavefunction, which is necessitated as the evolution is no longer unitary. The summation over the series (\ref{series}) in the present case is very easy, as the quantum evolution after the last collapse is what matters. The evolution from that state is either from level $\vert b \rangle$ or level $\vert c \rangle$. Suppose that the last $l$th collapse occurred to $\vert b \rangle$ level. Then
\begin{eqnarray}
Q_l |\Psi(0) \rangle &=& \int_{0}^{t} dt_1 ~e^{-\Gamma(t-t_1)} \Gamma ~U(t-t_1) \vert b \rangle ~P_l(t_1), \label{eq70}
\end{eqnarray}
where $P_l(t_1)$ denotes the probability that $l$ collapses have occurred in the interval 0 to $t_1$. This is given by
\begin{equation}
P_l(t_1) = \frac{(\Gamma t_1)^l}{l!} ~e^{-\Gamma t_1}. \label{eq71}
\end{equation}
The summation over $l$ now yields
\begin{eqnarray}
\sum_{l=1}^{\infty}Q_l ~|\Psi(0) \rangle &=& \Gamma \int_{0}^{t} dt_1 \left( 1- e^{-\Gamma t_1} \right) ~e^{-\Gamma(t-t_1)} \nonumber \\ && \times U(t-t_1) \vert b \rangle . \label{eq72} 
\end{eqnarray}
Here we have used the result:
\begin{equation}
\sum_{l=1}^{\infty} P_l(t_1) = 1 - e^{-\Gamma t_1}.
\end{equation}
One can write a similar expression if the last collapse occurred to level $\vert c \rangle$, by replacing $\vert b \rangle$ with $\vert c \rangle$. Combining these two possibilities with equal probabilities, we write the wavefunction as
\begin{eqnarray}
|\Psi(t) \rangle &=& \frac{1}{Z(t)} \Big[ e^{-\Gamma t} U(t) |\Psi(0) \rangle + {\Gamma \over 2} \int_0^t dt_1 \left( 1 - e^{-\Gamma t_1} \right) \nonumber \\ && \times e^{-\Gamma(t-t_1)} U(t-t_1) (\vert b \rangle + \vert c \rangle) \Big]. \label{eq74}
\end{eqnarray}
This is the final expression for the wavefunction in this model.

To compute it, we again resolve it in terms of instantaneous eigenfunctions:
\begin{equation}
\vert \Psi (t) \rangle = \displaystyle\sum_{k} W_k(t) e^{-i \int_{0}^{t} \lambda_k(t') dt'} \vert u_k(t) \rangle . \label{eq77}
\end{equation}
The coefficients $W_k(t)$ can be straightforwardly expressed in terms of $V_k(t)$s as
\begin{eqnarray}
W_k (t) &=& \frac{1}{Z(t)} \Big[ e^{-\Gamma t} ~V_k(t) + \frac{\Gamma}{2} \int_0^t dt_1 \left( e^{-\Gamma t_1} - e^{-\Gamma t} \right) \nonumber \\ && \times \displaystyle\sum_i V_i(t_1) e^{-i \left[ X_i (t_1) - X_k (t) \right]} \langle u_k(t) \vert u_i(t_1) \rangle \nonumber \\ && \times \left( \langle \Psi(0) \vert b \rangle + \langle \Psi(0) \vert c \rangle \right) \Big] , \label{eq79}
\end{eqnarray}
where $X_k (t) = \int_{0}^{t} \lambda_k(t') dt'$. 
We solve these equations numerically and compute the fidelities (\ref{eqfidelity}) and other relevant quantities in the storage and retrieval processes, as before, generalizing the coefficients $V_k$s in (\ref{eq30})-(\ref{eq32}) to the above $W_k(t)$s.

\section{Results and discussions}

We now present our results based on the equations developed in the last two sections. There are three relevant parameters which we take in the scaled forms of $r/\Omega_0$, $g_0/\Omega_0$ and $\Gamma/\Omega_0$. We have computed the variations of fidelities of the storage and retrieval processes with respect to all these parameters. The idea is to find the optimal parameters and develop an understanding of the dynamics of the transfer processes. A quantitative comparison with Berry's superadiabatic theory is presented in section 4.2 (see Table 1).

\subsection{Storage process}

\begin{figure}
\centering
     \resizebox{0.9\columnwidth}{!}{%
     \includegraphics{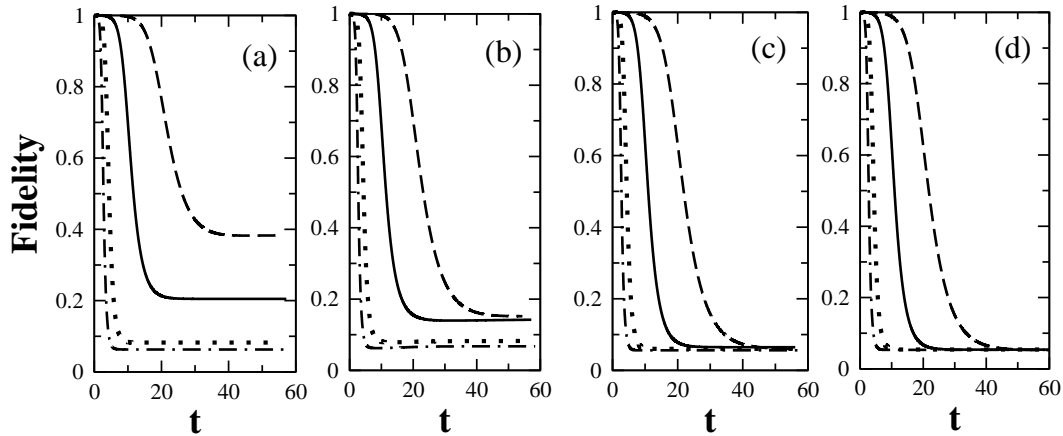}
}
\caption{Fidelity of the storage process versus time (in units of $\Omega_0^{-1}$) with a signal Rabi frequency of $g_0/\Omega_0 = 0.05$: (a) $\Gamma = 0$, (b) $\Gamma/\Omega_0 = 0.1$, (c) $\Gamma/\Omega_0 = 0.5$, and (d) $\Gamma/\Omega_0 = 1$. Each graph shows the effect of the variation of $r$: $r/\Omega_0 = 0.1$ (dashed), $r/\Omega_0 = 0.2$ (continuous), $r/\Omega_0 = 0.5$ (dotted), and $r/\Omega_0 = 0.8$ (dot-dashed).}
    \label{stog005}
\end{figure}

For the storage of a single photon from the cavity to the atomic memory with the control field of the form (\ref{eq50}), figure \ref{stog005} shows the fidelity (\ref{eqfidelity}) as a function of time (in units of $\Omega_0^{-1}$) for a relative signal field $g_0/\Omega_0 = 0.05$ without dissipation ($\Gamma = 0$) at (a), and with dissipation at (b) $\Gamma/\Omega_0 = 0.1$, (c) $\Gamma/\Omega_0 = 0.5$, and (d) $\Gamma/\Omega_0 = 1$, for different rates of variation $r$ (in units of $\Omega_0$) = 0.1, 0.2, 0.5 and 0.8 of the control field in each case. Figures \ref{stog01} and \ref{stog02} show the same set of results for higher signal field strengths, $g_0/\Omega_0 = 0.1$ and $0.2$, respectively. Note that the variation of the control field (\ref{eq50}) ends as it drops to zero at $t \sim 3/r$, and indeed one finds that there is no change in fidelity values after this time. 

\begin{figure}
\centering
     \resizebox{0.9\columnwidth}{!}{%
     \includegraphics{figure3.eps}
}
\caption{The same as figure \ref{stog005} but with a signal Rabi frequency of $g_0/\Omega_0 = 0.1$.}
    \label{stog01}
\end{figure}
\begin{figure}
\centering
    \resizebox{0.9\columnwidth}{!}{%
    \includegraphics{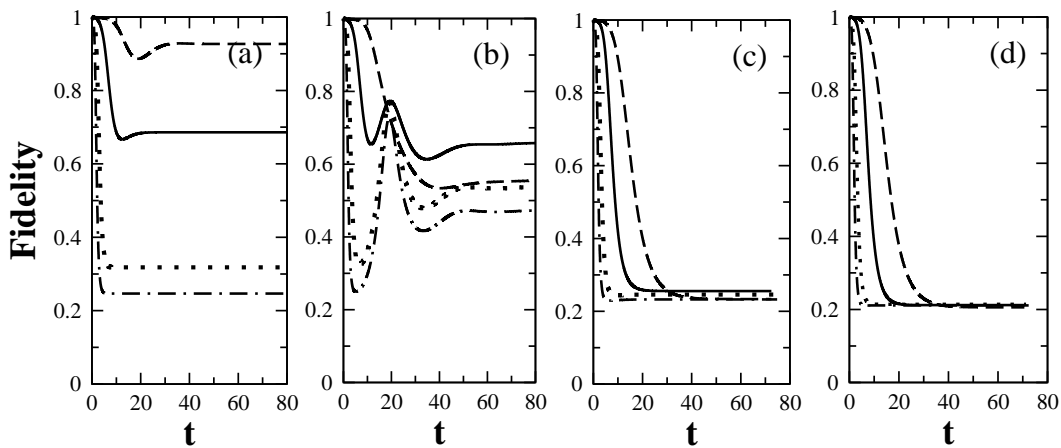}
}
\caption{The same as figure \ref{stog005} but with a signal Rabi frequency of $g_0/\Omega_0 = 0.2$.}
    \label{stog02}
\end{figure}

We first comment on the results without dissipation. In figure \ref{stog005}(a), one sees that the fidelities saturate to higher values as $r/\Omega_0$ decreases, for example, $F \sim 0.39$ for $r/\Omega_0=0.1$ and $F \sim 0.07$ for $r/\Omega_0=0.8$. Thus fidelity of storage is better with a slow variation of the control field, as expected from the adiabatic theory. The same is true for other values of $g_0/\Omega_0$, as shown in figures \ref{stog01}(a) and \ref{stog02}(a).

Next we examine how the fidelity depends on the signal strength. This variation is shown for a range of values of $g_0/\Omega_0$ from 0.05 to 0.2, in figures \ref{stog005}(a) to \ref{stog02}(a) for zero dissipation. It is seen that the fidelity increases with an increase of $g_0/\Omega_0$, leading to $F \sim 0.92$ for $r/\Omega_0 = 0.1$ at $g_0/\Omega_0 = 0.2$. This can be understood easily as the non-adiabatic perturbation (proportional to $\dot{\theta}$) becomes large when $\Omega_C(t) \rightarrow 0$, and it is larger, smaller the value of $g_0$.

We now examine the effect of dissipation. This is shown in figures \ref{stog005}-\ref{stog02} in panels (b), (c) and (d) with increasing values of $\Gamma/\Omega_0$. We see that with dissipation, the fidelities decrease, though the difference becomes marginal as $\Gamma$ increases. The variation with respect to $r/\Omega_0$ follows the same trend of decreasing fidelities with increasing $r/\Omega_0$, but is marginal for large $\Gamma$. Note that for a given system, $\Gamma$ is a fixed parameter; however, here the dissipation rate is scaled by $\Omega_0$ and hence its variation physically implies the inverse variation of the control laser power. 

\begin{figure}
\centering
     \resizebox{0.9\columnwidth}{!}{%
     \includegraphics{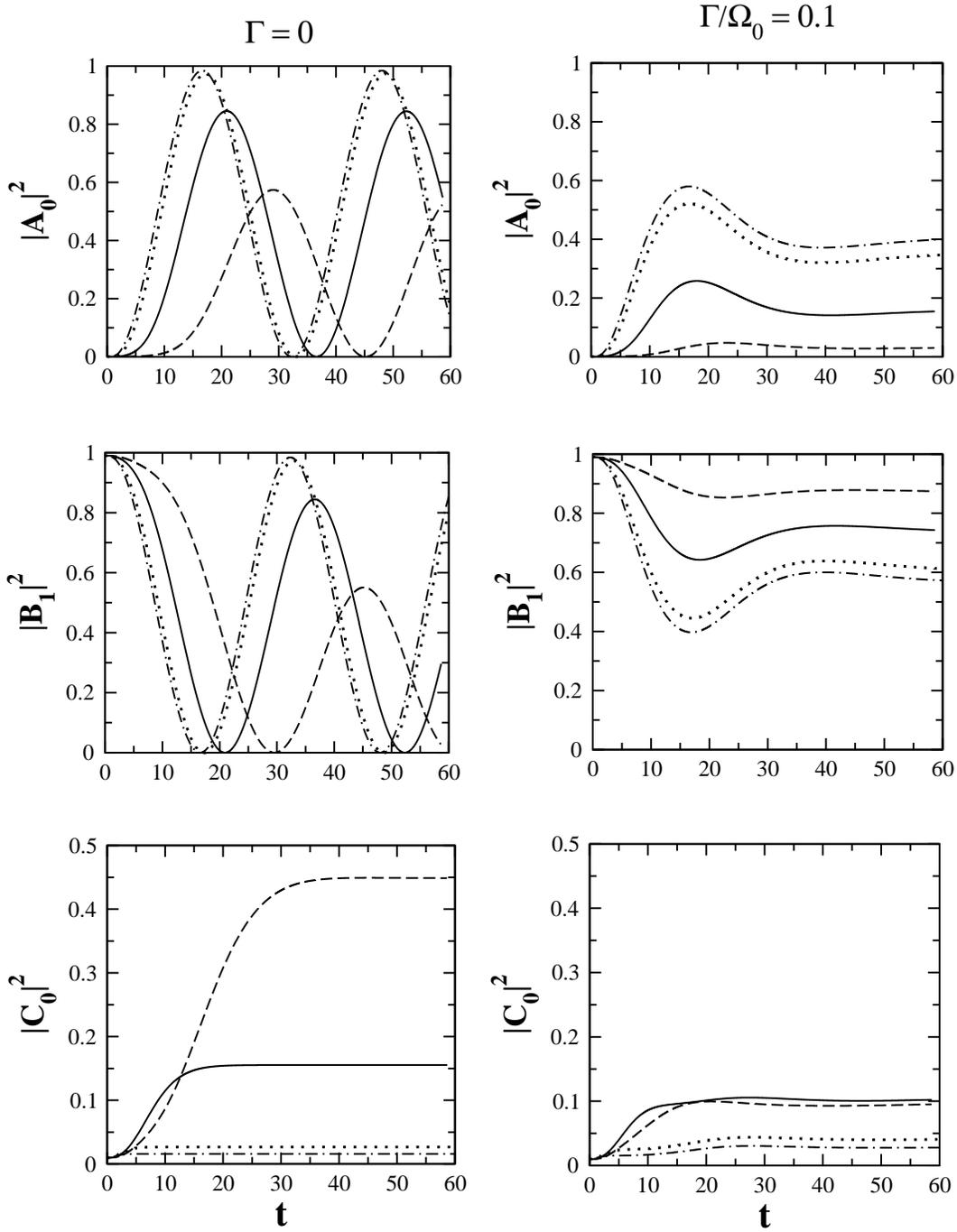}  
}
\caption{Plots of $\vert A_0(t) \vert ^2,~ \vert B_1(t) \vert ^2,~ \vert C_0(t) \vert ^2$ for the storage process, as a function of time (in units of $\Omega_0^{-1}$) with a signal Rabi frequency of $g_0/\Omega_0 = 0.1$ without dissipation, $\Gamma = 0$ (left-hand side), and with dissipation at $\Gamma/\Omega_0 = 0.1$ (right-hand side). Each graph shows the effect of the variation of $r$: $r/\Omega_0 = 0.1$ (dashed), $r/\Omega_0 = 0.2$ (continuous), $r/\Omega_0 = 0.5$ (dotted), and $r/\Omega_0 = 0.8$ (dot-dashed).}
    \label{stoABC}
\end{figure}

To give a detailed picture of the evolution of our solution for $\vert \Psi (t) \rangle$, the plots for the probability densities $\vert A_0(t) \vert ^2,~ \vert B_1(t) \vert ^2$ and $\vert C_0(t) \vert ^2$ of finding the system in the states $\vert a,0 \rangle$, $\vert b,1 \rangle$ and $\vert c,0 \rangle$, respectively, are shown in figure \ref{stoABC} in case of $g_0/\Omega_0 = 0.1$. On the left-hand side are the plots for no dissipation while on the right-hand side are those with dissipation at $\Gamma /\Omega_0 = 0.1$. In storage, as $\Omega_C \rightarrow 0$, the dark state approaches $\vert c,0 \rangle$. Thus $\vert C_0(t) \vert ^2$ should be large at the end of the process. However, without dissipation, this is marginally fulfilled for the smallest rate $r/\Omega_0 = 0.1$, and in the presence of dissipation, it is worse.  

\subsection{Retrieval process}

\begin{figure}
\centering
     \resizebox{0.9\columnwidth}{!}{%
     \includegraphics{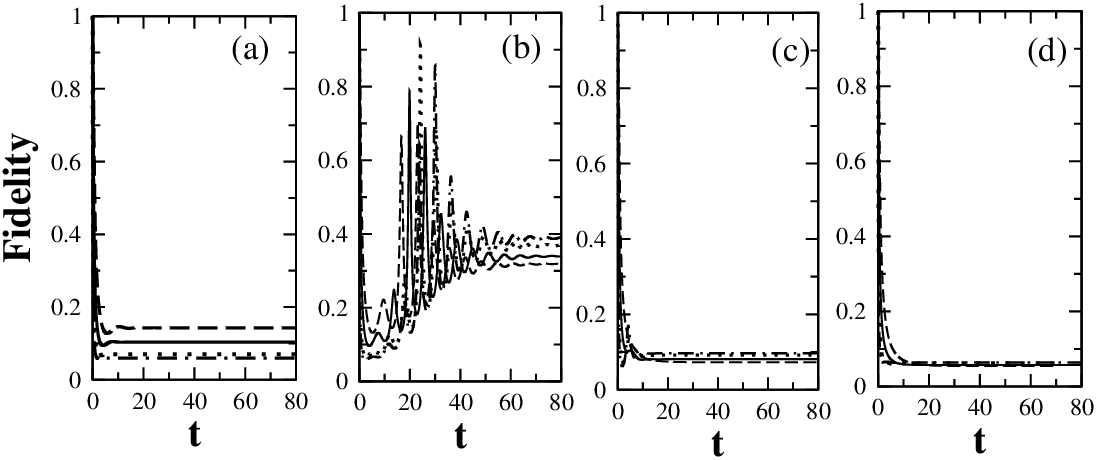} 
}
\caption{Fidelity of the retrieval process versus time (in units of $\Omega_0^{-1}$) with a signal Rabi frequency of $g_0/\Omega_0 = 0.05$: (a) $\Gamma = 0$, (b) $\Gamma/\Omega_0 = 0.1$, (c) $\Gamma/\Omega_0 = 0.5$, and (d) $\Gamma/\Omega_0 = 1$. Each graph shows the effect of the variation of $r$: $r/\Omega_0 = 0.1$ (dashed), $r/\Omega_0 = 0.2$ (continuous), $r/\Omega_0 = 0.5$ (dotted), and $r/\Omega_0 = 0.8$ (dot-dashed).}
    \label{retg005}
\end{figure}

\begin{figure}
\centering
     \resizebox{0.9\columnwidth}{!}{%
     \includegraphics{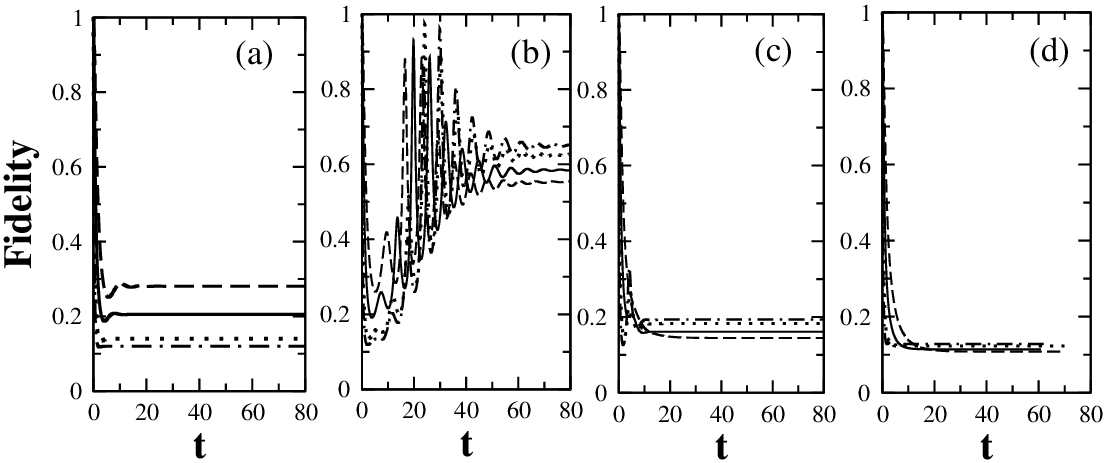} 
}
\caption{The same as figure \ref{retg005} but with a signal Rabi frequency of $g_0/\Omega_0 = 0.1$.}
    \label{retg01}
\end{figure}

\begin{figure}
\centering
     \resizebox{0.9\columnwidth}{!}{%
     \includegraphics{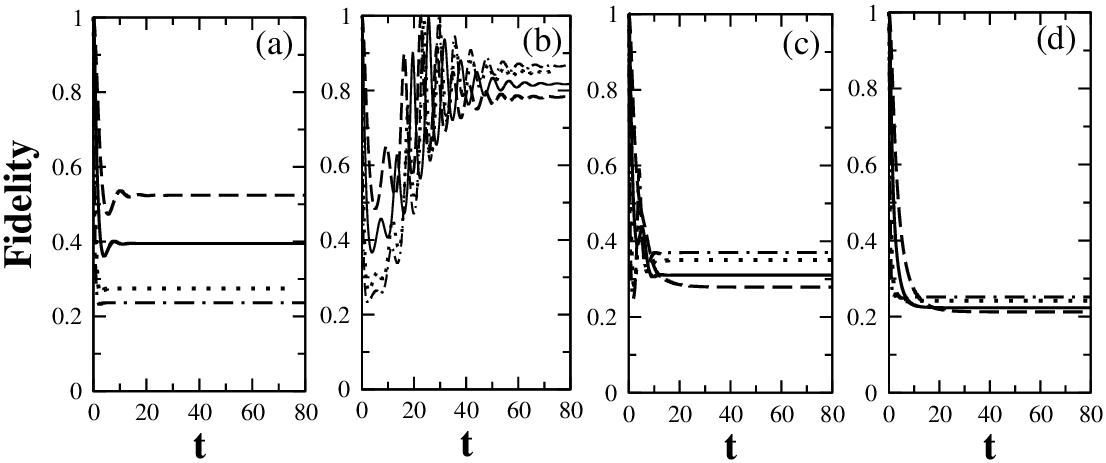} 
}
\caption{The same as figure \ref{retg005} but with a signal Rabi frequency of $g_0/\Omega_0 = 0.2$.}
    \label{retg02}
\end{figure}

Next we turn to the retrieval of the photon from atomic memory with the control field of the form (\ref{controlret}). Figure \ref{retg005} shows the fidelity (\ref{eqfidelity}) of the retrieval process as a function of time (in units of $\Omega_0^{-1}$) for a relative signal field $g_0/\Omega_0 = 0.05$ without dissipation ($\Gamma = 0$) at (a), and with dissipation at (b) $\Gamma/\Omega_0 = 0.1$, (c) $\Gamma/\Omega_0 = 0.5$, and (d) $\Gamma/\Omega_0 = 1$, for different rates of variation $r$ (in units of $\Omega_0$) = 0.1, 0.2, 0.5 and 0.8 of the control field in each case. Figures \ref{retg01} and \ref{retg02} show the same set of results for higher signal field strengths, $g_0/\Omega_0 = 0.1$ and $0.2$, respectively. We again note that the fidelity at a particular $r/\Omega_0$ saturates after a time $t \sim 3/r$ when the control field variation saturates to $\Omega_0$.


As before, we first comment on the results without dissipation. We note from figure \ref{retg005}(a) that the fidelities are much lower than that in the storage process in the range of values shown, even for rather small $r/\Omega_0$. They do decrease with increasing $r/\Omega_0$, but it is only a marginal effect. Thus the adiabaticity expectations are not quite fulfilled. This difference in the behavior of fidelity between storage and retrieval is easily understood by recognizing that for retrieval the non-adiabatic perturbation is the largest at the beginning of the procedure whereas for storage it is the largest toward the end of the procedure. 

As seen from figures \ref{retg005}(b)-(d), with dissipation the behavior is complex. For low dissipation rates (or high control power), the fidelity improves considerably but the behavior is non-monotonic. The best results are achieved with $\Gamma/\Omega_0 = 0.1$ and hereafter the results deteriorate with decreasing control power. Surprisingly, for retrieval we find that in the presence of dissipation, the general wisdom of adiabaticity is not followed; instead, the fidelity is better as $r/\Omega_0$ increases (except for very weak signal intensity, when $g_0/\Omega_0 \sim 0.01$, not shown here). 

In figures \ref{retg01} and \ref{retg02}, we mark the effect of the signal strength. It is seen from figures \ref{retg005}-\ref{retg02} that with an increase of the signal strength $g_0/\Omega_0$ from 0.05 to 0.2, the fidelity of retrieval improves, as for storage. For each signal strength, there is an optimum value of $\Gamma$ (or control power) at which the best fidelities are achieved. The behavior with respect to $r/\Omega_0$ shows the same unexpected trend.   

\begin{figure}
\centering
     \resizebox{0.9\columnwidth}{!}{%
     \includegraphics{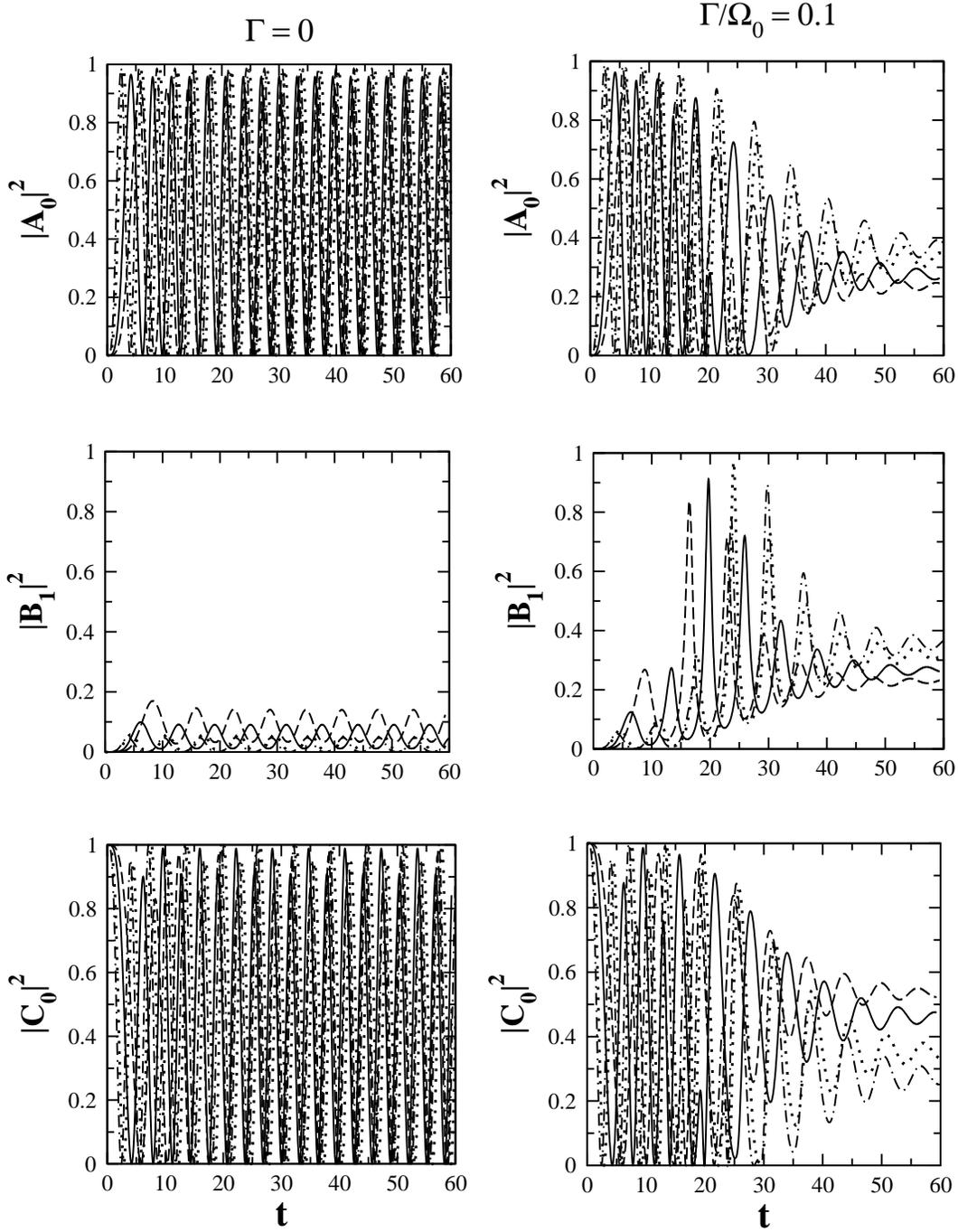} 
}
\caption{Plots of $\vert A_0(t) \vert ^2,~ \vert B_1(t) \vert ^2,~ \vert C_0(t) \vert ^2$ for the retrieval process, as a function of time (in units  of $\Omega_0^{-1}$) with a signal Rabi frequency of $g_0/\Omega_0 = 0.1$ without dissipation, $\Gamma = 0$ (left-hand side), and with dissipation at $\Gamma/\Omega_0 = 0.1$ (right-hand side). Each graph shows the effect of the variation of $r$: $r/\Omega_0 = 0.1$ (dashed), $r/\Omega_0 = 0.2$ (continuous), $r/\Omega_0 = 0.5$ (dotted), and $r/\Omega_0 = 0.8$ (dot-dashed).}
    \label{retABC}
\end{figure}

Again, to give a detailed picture of the evolution of our solution for $\vert \Psi (t) \rangle$, the plots for $\vert A_0(t) \vert ^2,~ \vert B_1(t) \vert ^2,~ \vert C_0(t) \vert ^2$, given in (\ref{eq30})-(\ref{eq32}) with the coefficients $V_k$s generalized to the $W_k(t)$s in (\ref{eq79}), are shown for the retrieval process in figure \ref{retABC} in case of $g_0/\Omega_0 = 0.1$. On the left-hand side are the plots for no dissipation while on the right-hand side are those with dissipation at $\Gamma /\Omega_0 = 0.1$. For retrieval, as $\Omega_C \rightarrow \Omega_0$, the dark state approaches $\vert b, 1 \rangle$ (for $\Omega_0 \gg g_0$). So, $\vert B_1(t) \vert ^2$ should be large at the end of the process. However, as we find in figure \ref{retABC}, $\vert B_1(t) \vert ^2$ is quite small without dissipation, but improves on inclusion of dissipation and it is better for a fast variation of the control field.  

It is finally noted that for slow rates of change of the control laser, the non-adiabatic corrections in the quantum evolution \textit{without dissipation} can be estimated from the superadiabatic transfer procedure given by Berry \cite{berry90}. This procedure has been applied to a $\Lambda$-system with $\Delta = 0$ by Elk \cite{elk95}, using its equivalence to a two-level description \cite{hioe90}. Applying the above result for the probability $P_{\infty}$ of transfer out of the dark state for the $\Omega_C(t)$-protocol (\ref{controlret}) for retrieval, we get 
\begin{equation}
P_{\infty} = \exp \left(- \frac{2 \pi (\sqrt{\Omega_0^2 + g_0^2}- \Omega_0)}{r} \right) .
\end{equation}
The fidelity is then given by $(1 - P_{\infty})$. We compare this result with our exact numerical computations in Table 1.

\begin{table}
\caption{Comparison of our steady-state retrieval fidelities $F$ for dissipationless cases ($\Gamma$ = 0) with Berry's superadiabatic transfer results using (41).}
\lineup
\begin{indented}
\item[]\begin{tabular}{@{}l|l|l|l|l|l|l}
\br
$g_0/\Omega_0 \rightarrow$ & \multicolumn{2} {c|}{0.05} & \multicolumn{2}{c|}{0.1} & \multicolumn{2} {c}{0.2} \\ \cline{2-7} 
$r/\Omega_0 \downarrow$ & Ours & Berry's & Ours & Berry's & Ours & Berry's \\
\mr
0.001 & 0.91 & 0.9996 & 0.99 & 1.0 & 0.999 & 1.0 \\
0.01 & 0.425 & 0.544 & 0.73 & 0.957 & 0.96 & 1.0 \\
0.1 & 0.14 & 0.076 & 0.28 & 0.269 & 0.53 & 0.712 \\
0.2 & 0.11 & 0.039 & 0.21 & 0.145 & 0.40 & 0.463 \\
0.5 & 0.08 & 0.016 & 0.135 & 0.061 & 0.28 & 0.220 \\
0.8 & 0.06 & 0.01 & 0.12 & 0.038 & 0.24 & 0.144 \\
\br
\end{tabular}
\end{indented}
\end{table}

As can be seen, our calculations are in good agreement with the superadiabatic theory, particularly for large $g_0/\Omega_0$. For small $g_0$ and large $r$, the non-adiabatic perturbations are strong and we do not expect the superadiabatic formula to give reliable results. In general, for small $r/\Omega_0$, the superadiabatic theory overestimates the fidelity, whereas at large $r/\Omega_0$, it underestimates the fidelity.

\section{Conclusions}

In this paper, we have addressed the core problem of adiabatic transfer driven by a control laser for a single three-level atom confined to a high-finesse optical cavity. We have studied the problem numerically by using instantaneous eigenstates of the system, and obtained results with and without dissipation. The dissipation in our case is included by allowing for spontaneous decay of the uppermost level of the $\Lambda$-configuration. Without the inclusion of dissipation, our results concur with the adiabatic idea of increased fidelity for slower variation of the control field. The dependence on the rate $r$ of variation is much weaker for the retrieval of the  photon than for its storage, especially for weak cavity-atom coupling. The fidelities of both the processes are better for higher coupling at any rate of variation of the control field. 

In the presence of dissipation, the fidelity of storage still follows the adiabatic property; however, it gets worse with higher dissipation, particularly at slow variations of the control field. For a given cavity-atom coupling, the fidelity of the storage process is better for higher ratios of control powers (Rabi frequency) to the spontaneous decay rate $\Gamma$. 

On the other hand, in the case of retrieval, the behavior with dissipation is non-monotonic. For an optimal control power $\Omega_0$, whose value depends on the spontaneous decay rate $\Gamma$ of the system concerned, the fidelity with dissipation is, in fact, better than that without dissipation, the best being at $\Gamma/\Omega_0 \approx 0.1$. Moreover, in the presence of dissipation, contrary to the adiabatic idea, retrieval fidelities are higher for faster rates of variation of the control field. 

In general, we conclude that an optimal dissipation aids the retrieval process but not the storage. Further, we find that the fidelity for storing is better, the stronger the control field and the slower the rate of its switching off. For the best fidelity of retrieval, however, there is an optimal power for the control laser dependent on the dissipation rate -- the faster the switching rate, the better is the retrieval. Our results lend themselves to experimental tests.

This quantum state transfer protocol, of course, has other practical limitations of implementation, e.g., residual atomic motion in the trap effectively reducing the atom-cavity coupling, imperfect preparation of the initial single-photon state in the cavity, etc. We have concentrated on the idealized model of a single atom trapped in a high-finesse cavity, which is relevant for studying the non-adiabatic issues discussed here in the presence of dissipation.

For various quantum information processing protocols, one needs to either couple the three-level atom to external photon channels or deal with several atoms. The full quantum mechanics of adiabatic transfer in such situations is highly complicated due to the enlarged Hilbert space of quantum states as well as the more complex dissipative processes. We feel that our present analysis is a useful input for tackling such problems, and some work is in progress in this direction. 

\ack
It is a pleasure to acknowledge illuminating discussions with Professor Michael Berry on his superadiabatic transfer formalism. We thank Professor Fabien Bretenaker for a critical reading of the manuscript. RG acknowledges support from an Indo-French Networking Project funded by the Department of Science and Technology, Government of India and the French Ministry of Foreign Affairs, and also from the Indo-French Centre for the Promotion of Advanced Research (IFCPAR/CEFIPRA). The work of JG was supported by the Council of Scientific and Industrial Research, India, and also by the University Grants Commission, India, under a special scheme of Capacity Build-Up in our University. 

\appendix
\section{Density-matrix approach for dissipative systems}
\setcounter{section}{1}

We describe here briefly the dissipation model, used in section 3, for the evolution of the density matrix $\rho (t)$. For a pure system,
\begin{eqnarray}
\rho (t) &=& \mathbf{U}(t) ~\rho (0) ~\mathbf{U^ \dagger}(t) \nonumber \\
&=& e^{-i \mathbf{H^ \times} t} ~\rho (0),
\end{eqnarray}
where $\mathbf{H^ \times} \mathbf{A} = \left[ \mathbf{H}, \mathbf{A} \right]$. To describe the spontaneous decay, we introduce the notation
\begin{equation}
 \mathbf{\zeta^ \times} ~\rho (t) = \mathbf{\zeta} ~\rho (t) ~\mathbf{\zeta^ \dagger}, 
\end{equation}
where $\mathbf{\zeta^ \times}$ can be $\mathbf{{\zeta_b} ^ \times}$ or $\mathbf{{\zeta_c} ^ \times}$. For example, 
\begin{equation}
 \mathbf{{\zeta_b} ^ \times} ~\rho (t) = \vert b \rangle \langle b \vert.
\end{equation}

Following the same approach as in the text, we include the possibilities of $1, 2, 3, \ldots, l, \ldots$ decays, with each of these possibilities weighted by their probability distribution. Then we have
\begin{equation}
\rho (t) = \displaystyle\sum_{l=0}^ {\infty} ~\rho_l (t),
\end{equation}
where
\begin{equation}
 \rho_0 (t) = e^{- \Gamma t} ~e^{-i \mathbf{H^ \times} t} ~\rho (0),
\end{equation}
and 
\begin{eqnarray}
\rho_l (t) = && \int_{0}^{t} \int_{0}^{t_1} \int_{0}^{t_2} ~\cdots \int_{0}^{t_{l-1}} e^{-\Gamma(t-t_1)} ~e^{-i \mathbf{H^ \times} (t-t_1)} ~\Gamma \nonumber \\ && \times dt_1 ~\mathbf{\zeta^ \times} e^{-\Gamma (t_1 - t_2)} ~e^{-i \mathbf{H^ \times} (t_1-t_2)} ~\Gamma ~dt_2 ~\mathbf{\zeta^ \times} \ldots \nonumber \\ && \times e^{-\Gamma (t_{l-1} - t_l)} ~e^{-i \mathbf{H^ \times} (t_{l-1}-t_l)} ~\Gamma ~dt_{l-1} ~\mathbf{\zeta^ \times} \nonumber \\ && \times e^{-\Gamma t_l} ~e^{-i \mathbf{H^ \times} t_l} ~\rho (0).
\end{eqnarray}
As is well known \cite{cb71}, this result is equivalent to the following evolution equation
\begin{eqnarray}
\frac{d \rho}{dt} &=& -i \left[ \mathbf{H^ \times} + i ~\Gamma \left( \mathbf{\zeta^ \times} - 1 \right) \right] \rho \nonumber \\
&=& - i ~\mathbf{H^ \times} \rho + \Gamma ~(\mathbf{W} \rho + \rho \mathbf{W^ \dagger}) + \Gamma ~\mathbf{W} \rho \mathbf{W^ \dagger}, \label{rho1}
\end{eqnarray}
where $\mathbf{W} = \mathbf{\zeta} - \mathbf{1}$. 

On the other hand, the wavefunction approach corresponds to the equation
\begin{equation}
i\frac{d \vert \Psi(t) \rangle}{dt} = \mathbf{H} ~\vert \Psi(t) \rangle + i \Gamma ( \mathbf{\zeta} - \mathbf{1} ) ~\vert \Psi(t) \rangle,
\end{equation}
which in turn yields for the density matrix
\begin{equation}
\frac{d \rho}{dt} = -i \mathbf{H^ \times} \rho + \Gamma ~(\mathbf{W} \rho + \rho \mathbf{W^ \dagger}) . \label{rho2}
\end{equation}
A comparison of (\ref{rho1}) and (\ref{rho2}) shows that the wavefunction approach agrees with the density matrix approach only up to first order in the `no-decay' operator $\mathbf{W}$.

\end{document}